# Contest in Multitasking: An Evidence from Chinese County Officials' Promotion Assessment

*By* YUANHAO ZHANG *

*Real-world observed contests often take the form of multi-task contests rather than single-task contests, and existing theories are insufficient to explain the incentive for extending the task dimension. This paper proposes a new effect of multi-task contests compared to single-tasking contests: the specialization effect (SE). By establishing a multi-task contest model with heterogeneous competitor costs, this paper shows that after expanding the new competition dimension, competitors will choose the dimension with greater relative comparative advantage rather than absolute advantage and pay more effort, which eventually leads to competitors choosing higher effort levels in both the original dimension and the extended dimension. The paper then uses staggered Difference-in-Difference (DID) method on China's county officers' promotion assessment from 2001 to 2022 as an entry point to discuss the empirical evidence for specialization effect. Through models and empirical studies, the specialization effect studied in this paper do exists in promotion assessments, and may also explain many other real-world scenarios, such as sports events, competition between corporate compensation and employee benefits and competition for R&D expenses.[1]*

* Yuanhao Zhang: University of Chicago, 1330 E 53$^{RD}$ St, Chicago, 60615 (e-mail: yuanhao@uchicago.edu).

# I. Motivation

Contests generally refer to situations in which two or more agents choose a specific level of effort to compete in order to win a limited prize. Many social phenomena can be abstracted into a contest model, such as sports events, university admissions, employee promotion competition, company research and development (R&D) competition, market product competition, rent-seeking competition, and arms races among countries. In the contest, whether the competitors (Contestants) win or not depends on their relative performance rather than absolute output, and the resulting strategic interaction of competitors is the core content of the theoretical study of the contest.

Real-world contests observed often appear in the form of multi-task contests rather than single-task contests , and existing theories are still insufficient to explain the motivation for expanding new dimensions of competition. For example: Why do some companies provide various company benefits (Company Benefits) instead of directly converting benefits into wages in salary design? Why are all kinds of self-enrollment and special examinations included in the university entrance examination? Why in the labor market, the overall quality of candidates is often more valued? Why in the Olympic trials, in addition to the selection of athletes for professional events, other physical fitness assessments are required for all athletes? Why is it necessary to measure the development level of economy, ecology, culture and other aspects when evaluating local governments? Although there may be multiple explanations for the above problems, this paper focuses on one of the motivations that is often overlooked: the specialization level enhancement effect.

Without considering Pareto optimality, the individual rational motivation of competitors to explore new dimensions of competition is clear. Assume that under a single-task contests, competitors are randomly assigned to "high" and "low" costs. When a competitor faces high costs, he has two options: continue to compete at high costs or expand a new mission. Obviously, if the cost of expanding new tasks is lower than the added value of expected revenue, it is in line with individual rationality for competitors to choose the latter. After choosing the latter, competitors will decide how much effort to allocate on different tasks according to their relative comparative advantages. Therefore, relative comparative advantage is the potential reason for the establishment of the specialization level enhancement effect. Therefore, some companies provide various corporate benefits: mining new corporate benefits costs almost nothing, but the company may have a comparative advantage in this task, thereby attracting more labor that prefers this benefit.

In the top-level design of the contests, mechanism designers will be able to use this relative comparative advantage to achieve the goal of improving the level of specialization. The meaning of the improvement of the specialization level has two layers: (1) the final total output in the expanded dimension is higher than the initial state; (2) more importantly, competitors staying in the original dimension will also improve their performance in the original task. The reason is that the expansion of dimensions makes people who did not have an advantage now have a comparative advantage, making the degree of competition more intense. Since the mechanism of comparative advantage determines the allocation of inputs across tasks, the mechanism designer can identify competitors with comparative advantages in a certain task through the design of multi-task contests to increase the degree of competition, although this division of a single task into multiple The practice of subtasks outside a certain threshold may reduce the total incentive. Therefore, whether the mechanism designer expands the new task dimension depends on the trade-off relationship between the gains brought by the improvement of the specialization level (or division of labor specialization) and the total incentive loss.

As an application of the effect proposed in this paper, this paper studies the evaluation of county promotion in China from 2001 to 2021. The core of the problem is: in the context of China's comprehensive deepening of reforms, how can the mechanism of official promotion effectively solve the principal-agent problem in local economic development, and fully consider the comprehensiveness and diversity of regional development while maximizing the efforts of officials? Sex (economic, ecological, cultural, etc.)?

In the past, various provinces in China measured the efforts of local officials based on county-level economic assessments and provided administrative incentives to the most outstanding local administrative officials, which effectively alleviated the principal-agent problem. County-level economic assessments often use GDP as a quantitative measure. After a simple classification of the counties under the jurisdiction of each province, they are sorted according to the GDP realization value of each county in each year, and the county-level administrative officials with the highest GDP are selected. Motivate. The county economic evaluation cycle is annual or semi-annual, and the specific means of administrative incentives are generally opportunities for promotion. This assessment system can also be called the "contests" mechanism, because it uses the actual GDP of other counties in that year as the benchmark.

However, with the successful completion of the "13th Five-Year Plan" and the implementation of the "14th Five-Year Plan", profound changes have also taken place in the evaluation methods of the county economy. In order to speed up the implementation of high-quality development by local governments at all levels, comprehensively deepen reforms, further accelerate the pace of county-level economic development, create a regional economy with both regional characteristics and complete functions, and give full play to the location advantages of each county-level place, each county-level place The government is constantly reforming and innovating in terms of administrative efficiency and administrative assessment system. Among them, a noteworthy changing trend is that the county economic assessment indicators have shifted from a single dimension to multiple dimensions. With the continuous implementation of the concept of sustainable development in various places, the economic assessment and evaluation methods in various places have gradually shifted from purely "GDP theory" to more multi-dimensional assessment, such as adding economic development, reform and opening up innovation, ecological civilization construction, social construction, people's livelihood and well-being, party The construction of the "five-type" government, subjective feelings and other evaluation indicators [1: such as Anhui Province, 2005], and through different index sorting methods, reward the counties that have achieved the best results in each index. At the same time, the assessment rewards have also changed from rewarding counties that ranked first in GDP to rewarding counties that ranked first in each indicator. However, compared with the previous single-dimensional economic assessment method, whether this change in the administrative assessment method can increase the efforts of local officials, thereby promoting the development of the county economy and bringing about improvements in the level of specialization of the county economy, etc. effect?

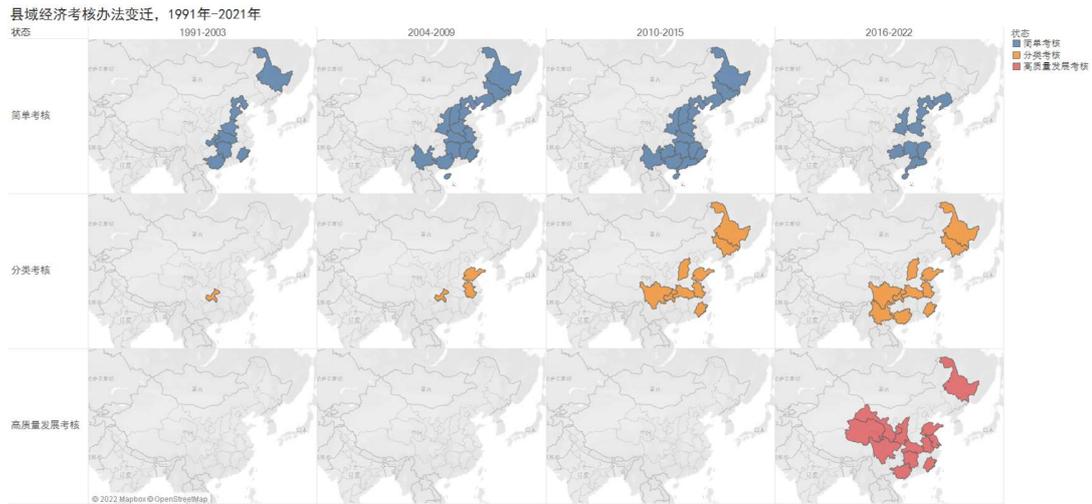

Note: (1) Simple assessment refers to the assessment method of "GDP-only" indicators (2) Simple classification assessment refers to the assessment method after dividing "first-class counties", "second-class counties" and "third-class counties" according to cost; The difficulty of different types of county assessments has been adjusted; (3) The assessment of high-quality development is based on the division of "first-class county", "second-class county" and "third-class county", and the assessment results after changing the assessment items and weights, and greatly increasing the ecological related indicators.

FIGURE1 CHANGES IN COUNTY ASSESSMENT METHODS, 1991-2021

The county economic assessment methods have undergone three rounds of changes:

(1) Before 1991-2010, simple assessments began in various places, mainly based on the county's GDP and other indicators to assess the county economy.

(2) From 2010 to 2015, some areas have gradually advanced from simple assessment to classified assessment. The common practice is to divide counties into "first-class counties", "second-class counties" and "third-class counties". In each category of counties, the most outstanding ones will be selected. However, during this period, the main goal of the classification assessment is still based on GDP growth.

(3) Beginning in 2018, the comprehensive evaluation system aimed at high-quality development has been launched, and the county-level evaluation has truly changed from a single-task evaluation to a multi-task evaluation.

In fact, the three methods of county economic assessment correspond to three different types of contests models: single-task single-prize contest, single-task multi-prize/multi-task single-prize contest, and multi-task contest. For single-task contests, there is currently the most relevant literature, and for multi-task contests, only a few articles have been studied so far. It is worth noting that a large part of this theoretical literature on multi-task contest, such as Lu and Fu (2009), does not actually reflect the heterogeneity of prizes or competitors in the task dimension, but only splits the prizes. Therefore, this type of literature still essentially studies single-task multi-prize contests. This paper introduces the heterogeneity of competitors in the task dimension by expanding the difference of the competitor's cost coefficient in the task dimension, to start the research on the multi-task contest.

TABLE1 TIME OF SIMPLE ASSESSMENT, CLASSIFIED ASSESSMENT AND HIGH-QUALITY DEVELOPMENT ASSESSMENT OF EACH PROVINCE

| province | Start a simple assessment time | Start sorting assessment time | Start the high-quality development assessment time | province | Start a simple assessment time | Start sorting assessment time | Start the high-quality development assessment time |
|---|---|---|---|---|---|---|---|
| Anhui | 2004 | 2008 | 2018 | Hunan | 2003 | NA | 2019 |
| Chongqing | 1999 | 2001 | NA | Jilin | 2005 | 2015 | NA |
| Fujian | 1994 | 2014 | 2019 | Jiangsu | NA | NA | 2018 |
| Gansu | NA | NA | 2019 | Jiangxi | 2007 | 2013 | 2019 |
| Guangdong | 2011 | NA | NA | Liaoning | 2006 | NA | NA |
| Guangxi | 2003 | 2016 | 2019 | Qinghai | NA | NA | 2018 |
| Guizhou | 2011 | 2013 | NA | Shandong | NA | 2005 | 2019 |
| Hainan | NA | 2008 | NA | Shanxi | 2006 | 2013 | NA |
| Hebei | 2003 | NA | 2019 | Shaanxi | 2004 | NA | 2019 |
| Henan | 1991 | NA | NA | Sichuan | NA | 2014 | 2019 |
| Heilongjiang | 2002 | 2014 | 2020 | Yunnan | 2004 | 2016 | NA |
| Hubei | 2003 | 2014 | 2019 | Zhejiang [1]* | NA | NA | NA |

Source: Compiled from provincial government websites. Among them, (1) simple assessment refers to the assessment method of "GDP-only" indicators, and (2) simple classification assessment refers to the assessment method after dividing "first-class counties", "second-class counties" and "third-class counties" according to cost; The difficulty of different types of county assessments has been adjusted; (3) The assessment of high-quality development is based on the division of "first-class county", "second-class county" and "third-class county", and the assessment results after changing the assessment items and weights, and greatly increasing the ecological related indicators.

From the time distribution, it can be seen that not all provinces have completely experienced the changes in the above three stages. Some provinces directly skipped the simple assessment stage or the classified assessment stage and entered the high-quality assessment stage (such as Jiangsu, Hunan, Hebei, Qinghai, Gansu, Yunnan). At the same time, the timeline of the evolution of various assessments is not uniform. As of 2021, 62.5% of the provinces have entered the stage of high-quality development assessment. Only Guangdong and Liaoning are still implementing simple assessment. my country's county economic assessment has entered a new stage.

TABLE 2 CHANGES IN EVALUATION METHODS: TAKE COUNTY ASSESSMENT IN ANHUI PROVINCE AS AN EXAMPLE

| Year of assessment | Classification situation | Categorization | Number of counties of various types | Assessment indicators | The number of metrics | Assessment methodology | Incentives | Send a text number |
|---|---|---|---|---|---|---|---|---|
| 1999 | Uncategorized | — | 61 | Development level + development vitality + development potential | 15[2] | Static 40% + Dynamic 60% | | Anhui Fa [1999] No. 6 |
| 2003 | Uncategorized | — | 61 | It is mainly based on GDP and per capita | 12 | Multi-index comprehensive evaluation method | Comprehensive top 10, dynamic top 10 report commendation + material rewards | Anhui Ban Fa [2004] No. 18 |
| 2008 | 3 categories | The cluster analysis method is divided into three categories considering the development conditions such as location and resources | 21+28+12 | Economic development + social development | 25 | Composite index, static 45% + dynamic 55% | The top 9 of the first category of counties, the top 9 of the second category of counties, and the top 4 of the third category of counties | Anhui Zheng [2008] No. 90 |
| 2012 | 4 categories | According to the main functional area planning, combined with the industrial foundation of various places and the assessment results in the past 3 years, it is divided into 4 categories | 16+13+19+14 | Economic development + transformation of development mode + resource and environmental protection + improvement of people's livelihood and social construction | 30 | Static 40% + Dynamic 60% | The top 7 first-class counties, the top 5 second-class counties, the top 6 third-class counties, and the top 5 fourth-class counties | Anhui Secretary [2012] No. 197 |
| 2018 | 4 categories | According to the main functional area planning, | 16+13+19+14 | Reduce the number of assessment indicators, | 15 | Static 40% + Dynamic 60% | The top 7 first-class counties, the top 5 second-class counties, the top 6 | Anhui Political Secretary |

[2] 12 in 2003 and 2004; 13 in 2005; In 2006, the indicators were divided into 26 comprehensive strong county indicators + 13 dynamic good county indicators

| | | | |
|---|---|---|---|
| combined with the industrial foundation of various places and the assessment results in the past 3 years, it is divided into 4 categories | improve the assessment indicators for ecological development, and especially divide "manufacturing strong counties" and "ecological strong counties" | third-class counties, and the top 5 fourth-class counties | [2018] No. 92 |

Source: According to the documents issued by the Anhui Provincial Government

The writing arrangement of this article is as follows: The second chapter of this article systematically sorts out the recent theoretical models of contest s, the changes of China's county assessment system and the model of the double-differences empirical method. Starting from the third chapter, this paper formally introduces two theoretical models, which are the single-task classification contest model and the multi-task contest model, and explains the multi-task contest model; the fourth chapter uses the multi-time point double difference method to analyze the Chinese county contest Model change empirically tests the degree of economic and ecological specialization. The fifth part discusses the main conclusions of the article, including the specialization level promotion effect, and the sixth part is the acknowledgment.

## II. Literature review

A lot of literature and empirical evidence have proved that contest s play an active role in solving the agent problem (Lazear and Rosen 1981) and mitigating moral hazard. However, the specialization effect newly proposed in this paper has not yet been proposed or tested in relevant literature. This paper mainly contributes to three literatures: the theoretical model of contests, the empirical test of county-level contest s in China, and the asymptotic double difference method.

### *A. Multi-task Contest Models and Related Theories*

The first literature is on theoretical modeling of multi-task contest models. The main contribution of this paper is to analyze the theory of two multi-task competitions under the condition of perfect information, when competitors are risk-neutral and heterogeneous in the cost function of different tasks. model, and proposed the specialization level promotion effect.

Theoretical model derivation has pointed out that in the case where competitors in the contest are risk-neutral and homogeneous, competitors have a linear cost function, and the total prize money of the contest is certain, single-task contests ("winner-take-all") The system design of the system has advantages over multi-task contests. For example, Lu and Fu (2009) proved that for a certain total prize money, dividing a single goal into multi-goal contests may reduce the overall motivation of participants, the smaller the number of participants, the greater the average prize money per participant More, the higher the total incentive of competitors; Arbatskaya and Mialon (2012) [13] also pointed out that under dynamic conditions, single-task contests still dominate multi-task contests.

This setting has a certain degree of flaws: According to the derivation results of the model, single-task contests are always dominant, and multi-task contests will not exist. In reality, the task dimension is expanded. Therefore, in recent years, more and more literature has focused on the following possible explanations for the fact that multi-task contests are superior to single-task contests. (1) The form of the competitor's cost function. Moldovanu, Benny and Sela (2001) [1] considered the contest model of two competitors, and first pointed out that when the cost function of the competitor's effort is in the form of a linear or concave function, it is more advantageous for the mechanism designer to choose a single-task contest ; and when the competitor's cost function is a convex function, a contest with 2 or more tasks may be more advantageous; Qualitative cost function cases, and find that in some cases contests possess numerical equilibrium solutions. (2) The degree of risk aversion of competitors. Fu Qiang, Xiruo Wang, and Zenan Wu(2021)[2], Fu Qiang, Xiruo Wang and Yuxuan Zhu(2021)[3] and other papers considered the introduction of uncertainty in the previous model, and the risk aversion of competitors In this case, the results show that multi-task contests dominate single-task contests when the risk aversion level of competitors reaches above a certain threshold. (3) Information asymmetry in the contest. Sigel (2014b) [8] proved that when there is information asymmetry between competitors, the equilibrium solution of the contest may change accordingly. On this basis, Olszewski, Wojciech and Siegel (2016) [9] gave a balanced estimation method. (4) The heterogeneity of rewards corresponding to each task. Jun Xiao (2016) summarized and analyzed the uniqueness of the equilibrium when there is sufficient information and the rewards corresponding to each task are different, and found that the equilibrium may not be unique in some cases. However, limited by the solution method, this theoretical literature has not yet given the form of the equilibrium solution.

TABLE 3 PRIZE SEQUENCES AND COST FUNCTION FORMS AS THEY APPEAR IN REFERENCES

|  | Award Function | Cost Function |
|---|---|---|
| Baye et al. (1996) | Single award | Differential Linear |
| Barut and Kovenock (1998) | Stochastic | Symmetric Linear |
| Clark and Riis (1998) | Homogeneous $v^k = v^{k+1}$ | Differential Linear |
| Bulow and Levin (2006) | Linear $v^k - v^{k+1} = \beta$ | Differential Linear |
| Siegel (2010) | Homogeneous $v^k = v^{k+1}$ | Stochastic |
| González-Díaz and Siegel (2013) | Linear $v^k - v^{k+1} = \beta$ | Non-linear |
| Jun Xiao(2016) | Quad $(v^k - v^{k+1}) - (v^{k+1} - v^{k+2}) = \beta$ | Non-linear |
| Jun Xiao(2016) | *Geometric* $v^k = \alpha v^{k+1}$ | Non-linear |

### B. Empirical Evidence of the county promotion contest in China

The second literature is an empirical test on the county promotion contest in China. The main contribution of this paper is to clearly distinguish the three stages of this contest: "simple assessment-simple classification assessment-high-quality classification assessment", and examine the differences in incentive levels, specialization degrees, etc. of different stages; and for the first time, it is proposed that the transition from a single task to a multi-task county-level assessment will help to improve the level of local specialization in a certain task, but on the other hand, the transition to multi-task may reduce the overall level of incentives. Furthermore, this paper analyzes the mechanism of the above conclusions.

Political contests are also known as promotion contests. Zhou Li'an (2007) first used the political contest mechanism at the county level to explain China's economic growth miracle since the reform and opening up. In the ten years since then, the research on using the contest mechanism to explain local administrative incentives has never stopped. In recent years, as the pace of my country's economic transformation and upgrading has accelerated, more and more scholars have noticed the changes in the contest model. Chen Zhao and Xu Tong (2011) pointed out that when the economic development reaches a certain stage, the society's dependence on public goods increases and information asymmetry differences weaken, the "GDP-only theory" performance appraisal method will no longer be the best in society. plan, the central government's governance of local governments has changed to a model of "competition for harmony". For this reason, they put forward the concept of "public satisfaction" in their article, and use this to demonstrate that "public satisfaction" is an important factor compared to "GDP". Better assessment indicators. On this basis, Wang Chunlei and Pan Jiayu (2020) believed that whether it is GDP or public satisfaction, single-dimensional assessment indicators cannot well explain that "competing for harmony " is an optimal way to maximize the welfare of local residents. governance model.

For this reason, they proposed that local public satisfaction is actually determined by multiple tasks such as medical care, culture, and the environment, and the weight of local assessment should be biased towards tasks with low information uncertainty.

When the multi-task contest system is applied to the assessment of Chinese officials in county areas, what impact will it have on the development of county areas? Can the multi-task contest system reach the overall social optimum compared with the single-task contest system? Many scholars have carried out empirical research around these issues. Jia Ruixue et al. (2017) found that under the framework of the multi-task contest, different goals have heterogeneous impacts on local borrowing behavior. Promotion incentives can curb the expansion of local government debt, but it will inhibit the local government's enthusiasm for economic development. Deng

Huihui and Zhao Jialing (2018) based on the panel data of 249 prefecture-level cities from 2006 to 2014, found that due to the existence of relative performance appraisal, local government officials would imitate and follow the same decisions as the "same Deng Xiaolan et al. (2019) took the scale of local debt as the research object, constructed a spatial econometric model to examine the local debt before and after the change in economic assessment, and found that competition for harmony can inhibit local Debt scale swelled. However, there is no literature to analyze whether the multi-task assessment system alleviates the principal-agent problem, thereby improving the total social welfare and improving the degree of specialization of the county economy.

### C. Insufficiency and improvement of Staggered DiD Method

The third branch of literature deals with improved estimates of the asymptotic difference-in-differences method. The Difference in Difference (DiD) method was first proposed by the medical scientist John Snow (1855) when he was studying the London epidemic and was introduced into economics by Obenauer and von der Nienburg (1915) when they were studying the effect of the minimum wage policy. The general difference-in-differences method requires that the treatment effect of the policy occurs at the same time point, thus constructing the difference-in-difference estimator . However the policies may not happen at the same time, just as the county-level assessment system changes that need to be studied in this paper are promulgated and implemented in each province in turn . In order to deal with this situation, progressive DiD (Staggered DiD) came into being. The commonly used progressive DiD is to construct the variable  to replace the variable of the standard double difference method $t \cdot treat_i trend_{it} t \cdot treat_i$

$$trend_{it} = \begin{cases} 0, & t < \bar{t}_i \\ t - \bar{t}_i, & t \geq \bar{t}_i \end{cases}$$

Among them,  represents the year when individual i was subjected to the intervention. Constructing is equivalent to forcing the treatment groups of variables at different intervention times to be aligned and grouped, and to uniformly compare the average treatment effect between them. The assumption behind this construction is that the treatment effects of policies do not vary systematically across individuals despite differences in the timing of the intervention. $\bar{t}_i trend_{it}$

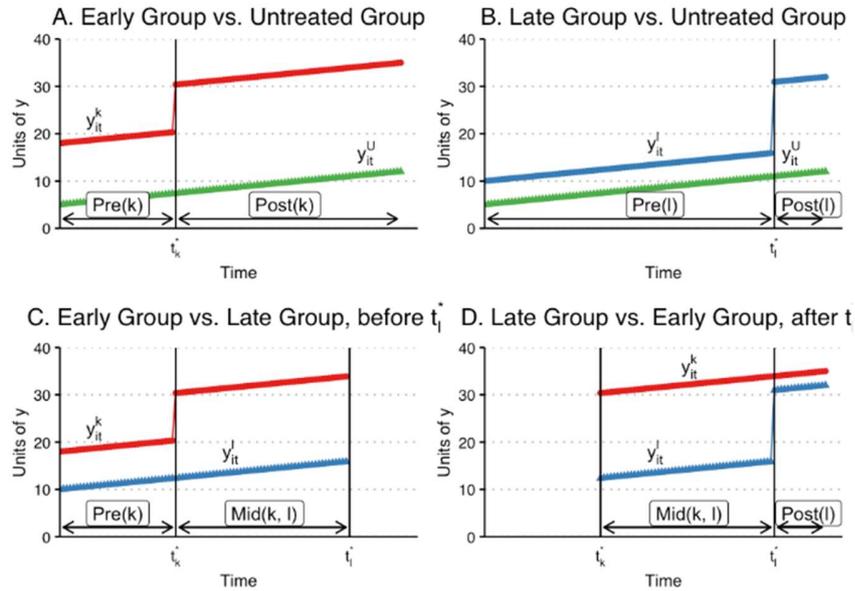

FIGURE 2 INFLUENCE OF MULTI-TEMPORAL DID ON THE WEIGHT OF THE MIDDLE GROUP (BAKER 2021).

Although 49% of DiD articles between 2010 and 2019 used progressive DiD (Baker 2021[20]), more and more articles found that the progressive DiD method itself may be flawed (Callaway and Anna 2020[21], Cengiz et.al 2019[22], Sun and Abraham 2020[23]). The biggest problem in the multi-time point progressive DiD is that the samples previously used as the treatment group (treat=1) may be used as the control group in the subsequent estimation. Therefore, treatment groups in the mid-term may have higher weights.

Therefore, this paper uses the Callaway and Anna statistic to reweight the progressive DiD and compares it with the general progressive DiD method.

## III. Theoretical Models

### A. Model Set-up

*Tasks.*— Suppose that n≥2 competitors compete on independent and $m \in \mathbb{Z}^+$ uncorrelated tasks, and each task selects the first-ranked competitor to be rewarded. Record the m tasks as $M_1, M_2, M_3 \ldots M_m$, and correspondingly, the reward corresponding to each task is recorded as $y_1 \geq y_2 \geq \cdots \geq y_m$.

*Reward Functions.*— Selecting the first competitor for each task can get some kind of reward. For example, rewards in the form of political promotion (such as direct promotion, transfer promotion, or increasing the potential probability of being promoted) in the county assessment contest. The reward function is in the form of homogeneity (Clark and Riis (1998), Siegel (2010), etc.), that is, the rewards for each task are the same, written as . At the same time, the total reward remains unchanged. In particular, when , the $y_1 = y_2 = \cdots = y_m \sum_{i=1}^{m} y_i = \bar{y} = 1$ $m = 1$ contest model degenerates into a single-task contest, and the reward function form is , $y_1 = \bar{y}$ which is the "winner takes all" situation.

*Contestants.*—n risk-neutral competitors participate in the contest , and their goal is to choose the appropriate effort to maximize their utility function. Assuming that competitors have the same form of utility function, after other competitors choose their efforts, the expected utility function of the kth competitor is as follows $e_k = [e_1 \ldots e_m]^T$

$$E(u_k; u_{-k}) = \sum_{i=1}^{m} [ \underbrace{P_k^i(e_{ik}) \times y_i}_{winning\ reward} - \underbrace{c_k^i(e_{ik})}_{effert\ cost} ]$$

Where

   $e_{ik}$: competitor 's effort on task $ki$

   $P_k^i(e_{ik})$: the prior probability of competitor winning the first place on task $ki$[3]

   $c_k(e_{ik})$: the cost of effort on the i-th task $e_{ik}$

---

[3] Based on Tullok(1980), $P_k^i$ is also called contest success function (CSF).

Although competitors are homogeneous in form of utility function, they show heterogeneity in marginal effort cost. The cost function of competitor k is represented by $c_k^i(x) = c_k^i x^2/2$, where the coefficient represents the marginal cost coefficient. In practical examples, the heterogeneity of the marginal cost coefficient represents individual advantages, such as the development advantages of the local economy (cultural advantages, ecological environment advantages, economic advantages, etc.) in the county $c_k^i$ contest. The larger means that the competitor k has a weaker advantage in the i-th task, and it will cost more to work hard on this task. The model does not set a hard constraint on the degree of effort, because when the marginal cost of effort reaches the marginal expected winning reward value, competitors will naturally stop continuing to work hard. Such a setting also shows that a competitor does not always have to fully allocate his energy, and he may potentially choose not to make any effort (if the cost coefficient is high enough, this choice is in line with personal rationality). $c_k^i$

Corresponding to the heterogeneity of the marginal effort cost, in order to ensure the effectiveness of the $c_k^i(x)$ contest mechanism, it is often necessary to compensate the weak, so as to increase the overall degree of competition and prevent "free-riding" behavior. For example, in a contest, the mechanism designer should group local units that have the same response to external shocks into the same category, so as to achieve a certain sense of "supporting the weak and suppressing the strong", and ultimately ensure the probability of everyone winning in each task Equal, reflected in the model as the heterogeneity on the winning probability function P^i, that is, ; you can also choose to win Probability does not intervene, and each local official chooses the task to work on according to his own advantages, that is, $P^i \neq P^j, \exists i, j \in \{1,2,\ldots,k\} P^i \neq P^j, \forall i, j \in \{1,2,\ldots,k\}$.

Define the winning probability as follows $P^i$

$$p_k^i = \begin{cases} \frac{w_{ik} e_{ik}}{\sum_{j=1}^n w_{ij} e_{ij}}, & e \neq 0 \\ \frac{w_{ik}}{\sum_{j=1}^n w_{ij}}, & e = 0 \end{cases}$$

Among them, is the degree of competition ambiguity. For simplicity, set . The larger means that the evaluation of the task is more beneficial to the local official k. Usually, the mechanism designer will assign a larger w_ik to the competitor k with a larger marginal effort cost to balance the achievement of the entire evaluation. for fair competition purposes. (Wang Chunlei,

Pan Jiayu (2020)) When , local officials will compete freely at this time, and the mechanism designer will not intervene in the final competition result. $w_{ik} \sum_i \mathbf{w_k} = 1, w_{ik} \in (0,1) w_{ik} i c_k^i w_{i1} = w_{i2} = \cdots = w_{ik}$

In particular, the case of county-level contests discussed in this article includes a simple classification assessment stage around 2014. This simple classification of contest competitors divided into N categories according to cost coefficients is essentially a mechanism design in which . At this time, the degree of ambiguity is a piecewise function of $w_k \neq 0 w_{ik} c_k^i$

$$w_{ik} = \begin{cases} a_1, & c_k^i \in (0, c_1) \\ a_2, & c_k^i \in [c_1, c_2) \\ \cdots \\ a_N, & c_k^i \in [c_{N-1}, 1) \end{cases}$$

The goal of constants is to select the degree of effort on each task to maximize the total expected utility $e_k = [e_1, e_2, \ldots, e_m]^T$

$$\max_{e_k} \sum_{i=1}^{m} [P_k^i(e_{ik}) \times y_i - c_k^i(e_{ik})]$$

*mechanism designer.—* In the model, it is assumed that the rewards are homogeneous and exogenously given, that is, , so the mechanism designer cannot change the size of the reward at will. At this time, the only thing the mechanism designer can change is the number of tasks. Therefore, the goal of the mechanism designer is to set an appropriate number of rewards or degree of divergence to maximize the overall level of effort. Taking the selection of the appropriate number of prizes as an example, let the overall effort be $y_i = y_j \mathrm{E}^*$

$$\max_{k \in \mathbb{Z}^+} \sum_{i=1}^{k} ||e_k|| = \max_{k \in \mathbb{Z}^+} E$$

### B. Suboptimality of Simple Classification, Single-Task Contests

First consider the contest model of a single task. The cost function of competitors shows heterogeneity in the single dimension . Mechanism designers have two options: Divide competitors into categories according to the size of the cost coeffici $N \ (N \leq n)$ ient (simple classification), or without any intervention in the entire $c_i$ contest (free competition). This section proves that simple classification contests have advantages over free competition contests, but simple classification contest s cannot maximize incentives. During the proof, the conditions that need to be satisfied to make the simple sorting contest optimal are explained at the same time. When optimal, the number of categories N must reach the level of the number of competitors , . In this case, each individual competitor will be classified into one class, so the optimal degree of dissimilarity should be a continuous function of $nN = nc_i$.

Given other constants' effort, under the condition of , contestant choose to optimize her expected utility $e \neq 0i$

$$\max_{e_k} \frac{w_k e_k}{\sum_{j=1}^{n} w_j e_j} \cdot y - \frac{c_k e_k^2}{2} \tag{1}$$

Then based on first order condition, we solve the equation that satisfies. Then let $e_k^* p_k = w_k e_k$, $p_k$ Can be viewed as the competitor's weight-adjusted effort。 The problem of choosing suitable $e_k^*$ level transform to suitable $p_k^*$ level

$$\frac{\partial E(u_k(e_k^*; e_{-k}))}{\partial e_k} = \frac{\partial E(u_k(p_k^*; p_{-k}))}{\partial p_k} w_k \tag{2}$$

$$= \frac{\sum_{j \neq k}^{n} p_j^*}{\left(\sum_{j=1}^{n} p_j^*\right)^2} y w_k - \frac{c_k p_k^*}{w_k} = 0$$

Notice that $w_k \in (0,1)$, so $\partial E(u_k; u_{-k})/\partial p_k = 0$. Under this equilibrium $E^{B*}$, $E^{B*} = \sum_j p_j$, therefore $\sum_{j \neq k}^{n} p_j^*$ can be written as $E^{B*} - p_k^*$. Plug the number in and we get

$$\frac{\partial E(u_k; u_{-k})}{\partial p_k} = \frac{E^{B*} - p_k^*}{(E^{B*})^2} y - \frac{c_k p_k^*}{w_k^2} = 0 \tag{3}$$

From this condition, we can obtain the condition that the weight-adjusted effort level satisfied by the competitors selected under the equilibrium level $p_k^*$

$$p_K^* = \frac{y E^{B*}}{(E^{B*}/w_k)^2 c_k + y} \tag{4}$$

Therefore, the degree of effort under the competitor's equilibrium condition is

$$e_k^* = \frac{y E^{B*}/w_k}{(E^{B*}/w_k)^2 c_k + y} \tag{5}$$

The degree of effort of competitors in equilibrium is related to the size of the reward y, the effort level adjusted by the total weight of all competitors, and the cost coefficient In fact, by further solving, we can get

$$\frac{\partial e_k^*}{\partial y} > 0 \tag{6}$$

$$\frac{\partial e_k^*}{\partial E^{B*}} < 0$$

$$\frac{\partial e_k^*}{\partial c_k} < 0$$

The practical implications of the above results are quite clear. When the value of the prize is higher, the effort level of all competitors is lower, and the cost coefficient of competitor k is lower, and the effort of the competitors is greater. Further, the competitor's winning probability is $P_k$

$$P_k^* = \frac{p_K^*}{E^{B*}} = \frac{y}{(E^{B*}/w_k)^2 c_k + y} \tag{7}$$

Next consider the overall optimal level of competition. The goal of the mechanism designer is to maximize the effort at the equilibrium level. Substituting the expression about the competitor's equilibrium effort level, we get $Ee_k^*$

$$\max_{w_k \in (0,1)} E = \max_{w_k} \sum_{k=1}^{n} \frac{y E^{B*}/w_k}{(E^{B*}/w_k)^2 c_k + y w_k} \tag{8}$$

The above-mentioned problem of finding the optimal degree of dissimilarity can be transformed into a problem of solving the optimal winning probability. According to the equilibrium optimal probability , , so $w_k P_k^* E^{B*}/w_k = \sqrt{(\frac{y}{P_k^*} - y)/c_k}$

$$\max_{w_k \in (0,1)} E \equiv \max_{P_K \in (0,1)} \sum_{k=1}^{n} P_k \sqrt{(\frac{y}{P_k} - y)/c_k} \tag{9}$$

$$s.t. \sum p_k = 1$$

The Lagrangian function of the above optimization problem is

$$\mathcal{L} = \sum_{k=1}^{n} \sqrt{(\frac{y}{P_k} - y)/c_k} + \lambda(1 - \sum_{k=1}^{n} p_k) \tag{10}$$

The first order condition is

$$\frac{\partial \mathcal{L}}{\partial P_k} = \frac{1 - 2P_K}{2\sqrt{c_k P_k(1 - P_K)/y}} - \lambda = 0 \tag{11}$$

Therefore, let $4(1 + \lambda^2 c_k/y) = a$, The winning probability of a competitor at equilibrium should satisfy the condition $aP_k^{*2} - aP_K + 1 = 0$

$$P_K^* = \pm\frac{\sqrt{a^2 - 4a}}{2a} + \frac{1}{2} \tag{12}$$

Obviously, the sum of winning probabilities at equilibrium should satisfy . The implicit solution of each competitor's effort obtained through this condition. The case where there is an explicit solution for competitor effort is discussed in the next section. $P_K^* = 1$

*Discussion.*—At this point, conclusions about the optimal classification can be obtained from the obtained formulas. From the expression in the expression of the optimal degree of ambiguity , it can be seen that after the task is $w_k = E_B^* \sqrt{c_k/(\frac{y}{P_k^*} - y)} w_k^*$ determined, in the equilibrium state corresponds to the level of one by one. This also confirms the role of the optimal discriminatory degree of "supporting the weak and restraining the strong". It can be seen that the optimal result of the final design of the tournament is to ensure that each competitor has the same prior probability of winning, so that whether the competitor can win the prize is completely determined by his own effort level. $w_k^* c_k$

Going from completely uncategorized tournaments to simple-category tournaments brought about an increase in overall effort levels, so the elimination of uncategorized tournaments by simple categories can be fully explained theoretically. In fact, by calculating the difference between the total effort levels of simple classification championships and unclassified championships, it can be found that the difference between the two is greater than 0, which shows that simple classification has indeed improved the overall effort level.

However, the simple classification tournaments in actual situations often only divide many local officials into three or four categories. In this case, it is obviously not the optimal degree of discrimination, because if corresponds to the level of $c_k$ one by one, and each When the economic development levels of the counties where the local officials are located are different, theoretically the adjustment weight for each county is different. Such a mechanism design obviously has practical operational difficulties. Therefore, when counties are divided into fixed $N < n$ categories, the optimal classification boundaries can also be discussed. By defining the sum of the Euclidean distances from the cost function of each local official to the nearest border, it is easy to prove that the optimal The bounds are the set of parameters {} that minimize the above sum. In fact, this classification method corresponds to the rationality of the clustering algorithm used in the grouping stage of the Chinese County Competition. $w_k^*$ $w_k^* c_1, c_2, \ldots, c_N \text{argmin}_{\{i\}} (\sum_{K=1}^{n} (c_K - c_i)^2) i$

### C. Optimality conditions for multi-task contests

For the multi-task tournament model, this paper discusses the effect of increasing the level of specialization: **whether in the extended** task dimension **or in the original** task **dimension, there are some competitors who invest more in a certain task than before.** The model in this section will point out that the strategy of allocating limited inputs according to competitor costs is still in line with the individual rationality of competitors, and competitors who stay on the previous dimension will also increase their input due to increased competition.

The multi-mission tournament considered in this article is a direct extension of the single-mission tournament model in the previous section, nwhere competitors compete in different dimensions. Let the overall level of effort on the task when equilibrium is reached $m$ $M_i$ as $E_i$,

$$\max_{e_{ik}} \sum_{i=1}^{m} \left( \frac{e_{ik}}{\sum_{j=1}^{n} e_{ij}} \cdot y_i - \frac{c_{ik} e_{ik}^2}{2} \right) \quad (13)$$

$$\text{s.t.} \sum_{i=1}^{m} e_{ik} = e_k \leq 1, k \in \{1,2,\ldots n\}$$

Here, competitors are subject to soft constraints on the total cost $e_k$ of effort. This is an important assumption about how much effort competitors need to make cross-task decisions about how much effort they put into each task in a way that is personally rational. Until then, however, the model can be simplified by the Lagrange method. Construct ka Lagrangian function with respect to competitors $\mathcal{L}(e_{ik})$, with Lagrangian multipliers $\lambda_k$ representing the shadow cost of the constraint

$$\mathcal{L}(e_{ik}) = \sum_{i=1}^{m} \left( \frac{e_{ik}}{E_i} \cdot y_i - \frac{c_{ik} e_{ik}^2}{2} \right) + \lambda_k \left( 1 - \sum_{i=1}^{m} e_{ik} \right) \quad (13)$$

$$= \sum_{i=1}^{m} \left( \frac{e_{ik}}{E_i} \cdot y_i - \frac{c_{ik} e_{ik}^2}{2} - \lambda_k e_{ik} \right) + \lambda_k$$

The Lagrange function reaches the optimal point when the following conditions are met. When deriving a particular $e_{ik}$ derivation, only the terms contained in the summation formula are retained, so the summation symbol in the formula $e_{ik}$ can be removed.

$$\frac{\partial E}{\partial e_{ik}} = 0 \Rightarrow \left(\frac{E_i - e_{ik}}{E_i^2}\right) y_i - c_{ik} e_{ik} - \lambda_k = 0, \quad i \in \{1,2,3,\dots m\} \quad (14)$$

$$\frac{\partial E}{\partial \lambda} = 0 \Rightarrow 1 - \sum_{i=1} e_{ik} = 0$$

Write an $e_{ik}^*$ expression under the optimal conditions based on the above conditions

$$e_{ik}^* = \frac{(y_i - \lambda_k E_i) E_i}{y_i + c_{ik} E_i^2} \quad (15)$$

The probability of winning in equilibrium is

$$P_{ik}^* = \frac{e_{ik}}{E_i} = \frac{(y_i - \lambda_k E_i)}{y_i + c_{ik} E_i^2} \quad (16)$$

The sum of the winning probabilities when the equilibrium is satisfied in a specific task dimension is equal to 1, therefore

$$\sum P_{ik}^* = 1$$

Based on the above conditions, the implicit solution form of the model can be finalized.

*Discussion.*—As in single-mission tournaments, the level of competition between competitors on multiple tasks at the optimal level depends on the total effort iallocated by all competitors on the task, $E_i$ their own competitive advantage in that task, and the level $c_{ik}$ of prize and shadow cost$(y_i - \lambda E_i)$. $e_{ik}^*$ Yes, $c_{ik}$ subtraction means that on multiple tasks, competitors tend to choose tasks that require minimal marginal cost. As a result, in the previous single-mission tournament, local officials who did not have an advantage in economic tasks could use their comparative advantages in ecological tasks to increase their chances of winning after the rules changed, and objectively tilted the level of effort towards ecological tasks.

Previous literature has shown that in cases where competitor risk is neutral and homogeneous, the change from a single prize to a multi-prize tournament reduces the overall effort level of the competitor, because in fact splitting a single reward into multiple points reduces the overall effort level of all (Lu and Fu (2009)). 。 However, this model is actually still a single-mission tournament model, because the M missions in this model are not really distinguished by tasks except in terms of prizes. Therefore, after distinguishing the competitors through the cost function, the total incentive reduction due to the dispersion of prizes pointed out in the previous literature and the total incentive improvement effect caused by the dimension expansion proposed in this paper compete with each other, and it is expected that the competition degree of the original task will increase.

In the case of two competitors, both models can solve the analytical solution to the above problem. To further simplify the model, prizes can be set for each task $y_i = 1$. When there is only one task, the cost factor of competitor 1 $c_{1k}$ is lower than that of competitor 2 $c_{2k}$, and the sum of the efforts of the original task dimensions is summed. When the extended task dimension has not yet been explored, assuming that the effort level of both competitors at this level is 0, the total effort level of the extended task dimension is naturally equal to 0.

In the case of a simple assessment of a single task, due to the moral hazard problem, competitors with cost advantages will not use the full degree of effort, but invest slightly lower than in the ideal situation; This situation has improved with the grouping of tasks, as evidenced by:

$$\frac{1}{2}\sqrt{c_1} + \frac{1}{2}\sqrt{c_2} \geq \frac{(c_1/c_2)^{1/4}}{(\sqrt{c_1} + \sqrt{c_2})} + \frac{(c_2/c_1)^{1/4}}{(\sqrt{c_1} + \sqrt{c_2})}$$

Intuitively, regrouping increases the level of competition among competitors, which motivates higher effort efforts.

Another solution is to extend the model to multiple dimensions; At this time, although there is no explicit grouping, from the level of effort, the effort of competitor 1 has reached a higher level than after grouping:

$$\frac{1}{2}\sqrt{c_{11}} + \frac{(c_{12}/c_{22})^{1/4}}{4(\sqrt{c_{22}} + \sqrt{c_{11}})} > \frac{1}{2}\sqrt{c_{11}}$$

The increment is caused by a higher level of dimensional competition. Although the total effort of Competitor 2 is missing in the original task dimension due to the shift of the effort level to another dimension, $\frac{(c_{12}/c_{22})^{1/4}}{(\sqrt{c_{22}} + \sqrt{c_{11}})} \frac{1}{2}\sqrt{c_2}$ there is a significant effort increase in the extended task dimension.

$$\frac{1}{2}\sqrt{c_{12}} + \frac{(c_{11}/c_{21})^{\frac{1}{4}}}{4(\sqrt{c_{11}} + \sqrt{c_{21}})} > 0$$

TABLE 4 ANALYTICAL SOLUTION OF THE TOTAL EFFORT LEVEL OF EACH TASK IN THE CASE OF A DOUBLE COMPETITOR

| | The original task dimension $E_1^*$ | Expand the task dimension $E_2^*$ |
|---|---|---|
| Simple assessment for a single task | $\dfrac{(c_1/c_2)^{1/4}}{(\sqrt{c_1}+\sqrt{c_2})} + \dfrac{(c_2/c_1)^{1/4}}{(\sqrt{c_1}+\sqrt{c_2})}$ | 0 |
| Multi-classification assessment of a single task (optimal ambiguity is selected by default) | $\dfrac{1}{2}\sqrt{c_1} + \dfrac{1}{2}\sqrt{c_2}$ | 0 |
| Multi-task extended assessment | $\dfrac{1}{2}\sqrt{c_{11}} + \dfrac{(c_{12}/c_{22})^{1/4}}{4(\sqrt{c_{22}}+\sqrt{c_{11}})}$ | $\dfrac{1}{2}\sqrt{c_{12}} + \dfrac{(c_{11}/c_{21})^{\frac{1}{4}}}{4(\sqrt{c_{11}}+\sqrt{c_{21}})}$ |

Note: It is initially assumed that competitor 1 has an advantage over competitor 2 in the initial task (task 1), that is, the advantage at the original task level does not change after the task is expanded, i.e.; $c_1 < c_2$ $c_{11} < c_{21}$ Suppose also that competitor 2 has an advantage over competitor 1 in scaling tasks, ie $c_{22} < c_{12}$. Finally, suppose that when the extended task (Task 2).

The above problems reveal the essence of the effect of increasing the level of specialization: in both the original dimension and the extended dimension, some competitors have increased their efforts.

## IV. Empirical Evidence

### A. Causal framework design

This paper intends to use the data of China's county assessment changes from 2001 to 2021, and use the multi-time progressive double difference (DiD) to estimate the causality, and test the effect of improving the specialization level in the two dimensions of ecology and economy. First, using the Rubin Causal Model (RCM) causal framework, an overview of the empirical model part of this paper is given. Three elements and four steps are emphasized in the RCM framework. The three elements refer to potential outcomes (potential outcomes), stability assumption (SUTVA, Stable Unit Treatment Value Assumption) and assignment mechanism (Assignment mechanism), and the four steps refer to modeling (model), identification (identify ), estimate and refute. Among them, modeling is to point out the independent variables, dependent variables, mediator variables, confounders, etc. of the causal model, and use these variables as the input of the model; identification refers to building a causal diagram and identifying all potential causal effects; Estimation refers to the estimation and interpretation of model parameters; refutation refers to performing robustness tests to verify that causal effects hold and avoid spurious causal relationships. In the previous part, the logical construction of the modeling part has been completed, and this section will formally use the causal language to express the modeling.

The main causality problem we are caring about is：

*What is the impact on county economy and ecological development from the transition from single-task competition to high-quality development multi-task competition?*

A Counterfactual equalization of the problem is ：

*If there is no transition from a single-task tournament to a multi-task tournament, how much will the county's economic and ecological development slow down?*

This is a potential consequence of the causal question that this paper is concerned with. In the above-mentioned causal problems, the individual of the potential outcome is each county in China from 2001 to 2021, and the intervention (treatment) of the potential outcome is that the provinces have introduced policies related to the transformation of the county-level assessment model. What this model actually studies is the Average Treatment Effect on The

Treated (ATT) of policy introduction. Let 0 and 1 represent pre-intervention and post-intervention, respectively, and use to represent the dependent variable of interest to individual i, then the average treatment effect on should be expressed by the following formula$Y_i$$Y_i$

$$\tau_{ATT} = \frac{1}{N_t} \sum_{i:D_i=1} (Y_i(1) - Y_i(0)),$$

If multitasking tournaments are expected to enhance specialization on extended tasks, it is inferred that a positive and significant incremental time trend for ecology-related indicators can be observed in the treatment group in ecological counties; whereas multitasking tournaments are expected to enhance specialization on original tasks If the level of globalization is higher, then it can be inferred that in the treatment group of economic counties, economic related indicators can be observed to have a positive and significant incremental time trend.

TABLE 5 TREND OBSERVATIONS FOR TREATMENT AND NON-TREATMENT GROUPS

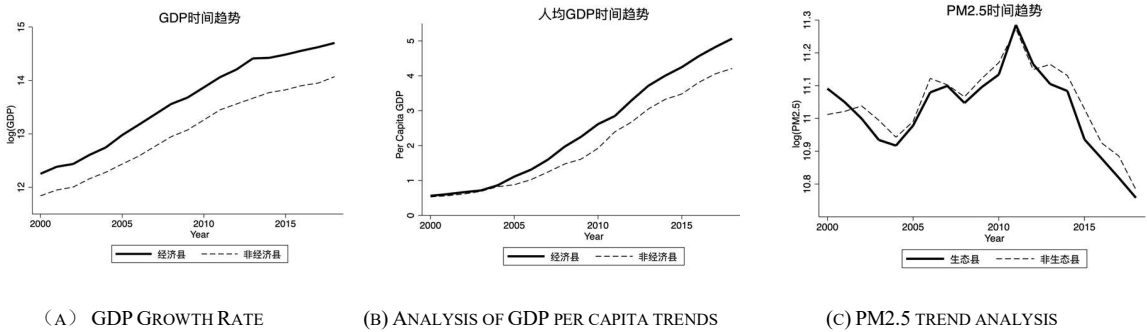

（A） GDP GROWTH RATE  (B) ANALYSIS OF GDP PER CAPITA TRENDS  (C) PM2.5 TREND ANALYSIS

## B. Data Sources

This paper collates and collects the CSMAR database, CNRDS database, CEI database, Dalhousie University, the government of the People's Republic of China and the websites of various local administrative units, from 2001 to 2021 for nearly 20 years variables related to the county's economy, environment, finance, and employment; And sorted out and collected relevant documents, newspaper news, etc. from the central or local government, sorted out the time for the implementation of classification assessment in various places, and the data sources are as follows.

TABLE 6 NAMES AND SOURCES OF DATA USED IN THIS ARTICLE

| Data name | Data sources |
|---|---|
| Fiscal revenue and expenditure of each county (sub-project) | CSMAR database |
| Income of urban and rural residents in each county | CSMAR database |
| Employees and wages of urban units in each county | CSMAR database |
| Statistics of employees in each county | CSMAR database |
| GDP by prefecture (by industry) | CSMAR database |
| GDP and index of each county | CSMAR database |
| Population, number of households and urbanization rate of each county | CSMAR database |
| Administrative divisions of counties | CSMAR database |
| Changes in administrative divisions at or above the county level | CSMAR database |
| City elevation | CNRDS database |
| PM2.5 data in county-level areas in China | Atmospheric Composition Analysis Group, Dalhousie University |
| The administrative division code of the county | Ministry of Civil Affairs of the People's Republic of China |
| County GDP | CEI database |
| County assessment data | The people's governments of all provinces (autonomous regions). |
| List of Counties in National Poverty in 1994 | Leading Group for Poverty Alleviation and Development of the State Council of the People's Republic of China |

## C. Identification

The task of this section is to intuitively give the basis for the increase in the level of economic/ecological specialization of the county before and after the policy, after a simple processing of the raw data.

The history of changes in China's county economic assessment spans more than 20 years, during which it experienced both the upsurge of "withdrawing counties and establishing cities" around 1996 and "withdrawing cities and establishing counties" in the early 21st century. In order to ensure the uniformity of each local sample and continuously observe the same sample, this paper matches and deletes counties that have changed administrative units. After processing, a total of 3 4015 observations were made from 1795 samples.

TABLE 7 CHANGES DURING THE SAMPLE OBSERVATION PERIOD

| Change type | Frequency statistics | percentage | Cumulative percentage |
|---|---|---|---|
| Withdraw from districts | 52 | 13.07 | 13.07 |
| Districts are withdrawn and counties are established | 2 | 0.5 | 13.57 |
| Districts shall be withdrawn to establish county-level cities | 1 | 0.25 | 13.82 |
| Withdraw districts and establish cities | 1 | 0.25 | 14.07 |

| | | | |
|---|---:|---:|---:|
| County-level cities are abolished and districts are established | 80 | 20.1 | 34.17 |
| Counties are abolished and districts are established | 143 | 35.93 | 70.1 |
| Counties were abolished and counties were established | 1 | 0.25 | 70.35 |
| Counties are abolished and county-level cities are established | 39 | 9.8 | 80.15 |
| Counties are abolished and cities are established | 2 | 0.5 | 80.65 |
| Counties are abolished and autonomous counties are established | 1 | 0.25 | 80.9 |
| The city is abolished and districted | 11 | 2.76 | 83.67 |
| Towns and districts were withdrawn | 4 | 1.01 | 84.67 |
| Towns are abolished and county-level cities are established | 1 | 0.25 | 84.92 |
| Renamed | 60 | 15.08 | 100 |
| total | 398 | 100 | 100 |

Further examine the core distribution rule in the article, that is, the counties that are classified as economic counties or ecological counties. The main concern here is the balance of data under allocation rules. An unbalanced set of data can affect the model's estimation effect, which can lead to bias. Ideally, therefore, there would be no large differences in the number of individuals between the intervention groups and between the intervention and control groups.

Prior to the policy intervention, there was no distinction between economic counties and ecological counties, as there was only a single-mission tournament format. Therefore, the balance test here mainly considers the number of economic counties or ecological counties after policy intervention. After policy interventions, 390 districts were classified as ecological counties and 4,83 as economic counties. It can be seen that the number of economic counties and ecological counties is roughly the same (483, 390), and there are no counties that are divided into both economic and ecological counties.

TABLE 8 BALANCE TEST FOR DISTRIBUTION RULES: CROSSTAB COUNT

| | | Eco County | | total |
|---|---|---|---|---|
| | | not | be | |
| Economic counties | not | 922 | 390 | 1312 |
| | be | 483 | 0 | 483 |
| total | | 1405 | 390 | 1795 |

**Next, descriptive statistics are performed on the core covariates (control variables) used in this article.** This paper intends to use double fixed effects to control factors that are not temporal or not geographically variable, so the variables that need additional control are mainly variables with time-varying characteristics and potentially potentially affecting the effects of policy treatment. At the same time, in order to avoid the potential impact of extreme values of

covariates on the estimation of policy treatment effects, this paper performs a 1% Wasor tail reduction treatment on the basis of the original estimate.

TABLE 9: DESCRIPTIVE STATISTICS FOR CORE COVARIATES

| | Number | mean | standard deviation | p1 | P99 | T value |
|---|---|---|---|---|---|---|
| Whether poor counties | 25143 | 0.240743 | 0.427543 | 0 | 1 | 89.28578 |
| Whether county boundaries | 25143 | 0.520463 | 0.499591 | 0 | 1 | 165.19 |
| Number of townships (units). | 21112 | 16.23551 | 8.471244 | 4 | 48 | 278.4731 |
| Number of fixed telephone households (households). | 22822 | 76770.04 | 98711.82 | 1500 | 452500 | 117.4896 |
| Registered population (person). | 23012 | 50.68562 | 35.57064 | 3.14 | 159 | 216.1575 |
| Number of employees in the secondary industry (person). | 20417 | 47584.23 | 73548.62 | 1000 | 352069 | 92.44524 |
| The added value of the primary industry (yuan). | 23012 | 166477 | 168802.3 | 4900 | 801808 | 149.6074 |
| The added value of the secondary industry (yuan). | 23011 | 630318.2 | 1219552 | 2300 | 5613902 | 78.40203 |
| Financial budget revenue (million yuan). | 22996 | 76719 | 175347.6 | 400 | 772850 | 66.3482 |
| Financial budget expenditure (million yuan). | 22996 | 165522.8 | 210535.1 | 5800 | 950576 | 119.2229 |
| Total resident savings deposits (yuan). | 22844 | 741367.7 | 1283056 | 6100 | 5839200 | 87.33215 |
| Loan balance of financial institutions at the end of the year (yuan). | 22861 | 761540.2 | 1990797 | 4400 | 8688500 | 57.83803 |
| Total grain production (tonnes). | 19665 | 243514.9 | 274724.2 | 3700 | 1250879 | 124.3013 |
| Number of industrial enterprises above designated size | 22736 | 120.721 | 347.4738 | 1 | 1134 | 52.38629 |
| Investment in fixed assets (yuan). | 13377 | 754716.7 | 1257404 | 1800 | 6131858 | 69.42063 |
| Number of general secondary schools (persons). | 23009 | 28429.35 | 23395.98 | 700 | 109400 | 184.3209 |
| Number of Primary School Students (Person) | 23008 | 40856.15 | 35081.13 | 2700 | 171800 | 176.6539 |
| Number of health care beds (beds) | 20498 | 1323.343 | 1238.244 | 100 | 5966 | 153.0107 |
| Number of social welfare units (pcs) | 22672 | 16.17109 | 91.22447 | 0 | 80 | 26.6915 |

**Then, descriptive statistics are performed on the main outcome variables in the model and the control variables for heterogeneity to be verified, and whether they are significant at the level are listed in the subscript (see the table below).** $p = 0.05$ Define binary processing variables and respectively to indicate whether a county is classified as an economic $treatment\_economy_{it}$ $treatment\_ecology_{it}$/ecological county. For example, when a county is classified as an ecological county, and before that. In this way, the whole sample can be classified according to the above two variables, and it can be seen that from treatment_ecology=0 to $treatment\_ecology_{it} = 1$ $treatment\_ecology_{it} = 0$ treatment_ecology=1, several indicators related to environmental pollution have been significantly reduced (). $\Delta pm\_25\_mean = -13.38, \Delta lpm25 = -0.03$ It means that the samples after being classified as ecological counties perform better in environmental pollution than the samples that are not classified as ecological counties; Then, from t reatment_economy=0 to treatment_economy=1 grouping, several indicators of economic development improved significantly ($\Delta lgdp = 0.59, \Delta pergdp = 0.5, \Delta urbanrate = 0.08$)。 Finally, as a placebo test, the descriptive statistics retained the economic development index of the sample when the treatment_ecology changed from 0 to $\Delta lgdp = -1.03^*, \Delta pergdp = -0.77^*, \Delta urbanrate = -0.011$ Ecological development index of the sample when the treatment_economy changes from 0 to 1 (. In addition, in order to prevent the expectation of 0 for the sample size condition of the policy treatment group under any control variable, which undermines the effectiveness of the double difference method, the basis for the overlap ($\Delta pm\_25\_mean = 5.11^*, \Delta lpm25 = 0.02$)Overlap) of several important control variables in this paper is also shown in the table.

Descriptive statistics intuitively suggest that the division of economic counties/ecological counties by multi-mission tournaments may affect the degree of economic/ecological specialization of the county, but there is no causal basis. From the observed results, it is also possible that the places classified as economic counties/ecological counties have strong economic development strength or ecological potential, and the economic or environmental indicators of these counties are already good in advance. This possibility manifests itself in the $C_i^k$ relative exogenous nature of the cost factor in the model. Therefore, in the next part, consider using a causal inference model for further analysis.

TABLE 10: SAMPLE DESCRIPTIVE STATISTICS FOR RESULT VARIABLES AND TEST FOR HETEROGENEITY TEST COVARIATES

| Table A: Result variables | Eco County | | | | Economic counties | | | | difference (conditional difference). | |
|---|---|---|---|---|---|---|---|---|---|---|
| | Control group | | Processing groups | | Control group | | Processing groups | | | |
| | average value | standard deviation | average value | standard deviation | average value | standard deviation | average value | standard deviation | Eco group | Economic group |
| g Logarithmic mean of g dp | 13.42a | 1.2 | 12.39b | 1.3 | 13.03a | 1.3 | 13.62b | 1.2 | 1.03 | 0.59 |
| Average GDP per capita | 2.21a | 2.3 | 1.44b | 1.3 | 1.88a | 2.1 | 2.38b | 2.1 | 0.77 | 0.50 |
| pm2.5 mean | 45.36a | 21.3 | 31.98b | 20 | 41.08a | 21 | 46.19b | 23.2 | 13.38 | 5.11 |
| pm2.5 logarithmic mean | 11.05a | 0.6 | 11.02b | 0.7 | 11.04a | 0.6 | 11.06a | 0.7 | 0.03 | 0.02 |
| Table B: 2,000 years of covariates | | | | | | | | | | |
| Level of urbanization | 0.18a | 0.1 | 0.17a | 0.1 | 0.16a | 0.1 | .24b | 0.2 | 0.01 | 0.08 |
| Proportion of poor counties | 20.9%a | 2.2% | 39.7%b | 1.5% | 29.0%a | 2.8% | 13.9%b | 2.7% | 18.80% | 25.00% |
| The proportion of the county boundaries | 47.0%a | 4.3% | 57.4%b | 3.7% | 52.1%a | 5.4% | 41.8%b | 4.3% | 10.40% | 49.30% |
| Added value of the secondary industry | 45.2%a | 1.7% | 51.7%b | 1.9% | 46.4%a | 5.2% | 57.3%b | 2.1% | 7.50% | 10.9% |
| Budget expenditure | 139028 | 1543 | 125527 | 1695 | 125420a | 4520 | 186594b | 5510 | 13501 | 61174 |
| Table C: Differences in other county indicator conditions after controlling for variables in the preceding table | | | | | | | | | | |
| Percentage of education | 90.9%a | 2.2% | 92.7%a | 1.7% | 91.0%a | 1.6% | 91.9%a | 1.6% | 0.1% | 0.9% |

| Percentage of savings balance | 20.7%a | 1.3% | 20.4%a | 1.5% | 21.8%a | 0.1% | 21.5%a | 0.1% | 0.3% | 0.3% |
|---|---|---|---|---|---|---|---|---|---|---|
| Proportion of large-scale enterprises | 48.2%a | 0.7% | 51.7%a | 0.2% | 53.7%a | 0.3% | 50.2%a | 0.2% | 3.5% | 3.5% |

Note: (1) The subscripts a and b show a bilateral equivalence test for means under the assumption of homogeneity of variance (p< 0.05), When both a and b appear in the same row, it means that the mean is statistically significant. (2) Conditional difference refers to the difference between the treatment group and the control group after controlling the variables in Table B.

### D. Estimate: Multi-temporal asymptotic DiD and PSM-DiD

The empirical part of this paper mainly studies the changes **that have occurred before** and after ecological development and **economic development** from single-task to multi-task county assessment. In this paper, a double difference (DID) model is constructed by using proxy variables on environmental development (PM 2.5 emissions) and economic development (GDP-related indicators) as dependent variables, and the interaction terms between dummy variables and time for high-quality assessment or classification assessment as the multiplier estimation measure. In building the model, the following key factors are taken into account:

1、Controlling fixed effects: Fixed effects include temporal fixed effects and individual fixed effects. The individual fixed effect is divided into two dimensions, namely the province dimension and the county dimension.

   （1） **The rationality of controlling the fixed effect of provinces:** As mentioned in the introduction, the promotion quota of county administrative officials is generally divided among the provincial administrative units, and the assessment content of the county competition in each province is different, and the actual impact of policies in different provinces may also be different. These questions ultimately influence the model's estimation of policy effects. Therefore, in the model, it is necessary to control the provincial fixed effect.

   （2） **Rationality of controlling the fixed effect of counties**: Different counties have different levels of economic/ecological development before receiving intervention, and may be affected by relatively exogenous variables such as geographical location and climatic conditions that are not strongly related to policies (for example, because

the location advantages of coastal cities due to geographical conditions will promote local economic development, but whether they are adjacent to the sea is not necessarily related to whether they are classified as economic counties, Therefore, when this advantage is not controlled by the model, the economic growth brought by some geographical advantages is absorbed by the coefficients of the intervention variable, which brings an upward estimation bias of the treatment effect). Controlling the county fixed effect can alleviate this part of the model bias, and controlling the county fixed effect is equivalent to controlling the above-mentioned provincial fixed effect.

（3） **Rationality for controlling temporal fixed effects:** The multiple estimators of the double difference estimation model estimate the temporal trend of policy effects, but do not include macroscopic influencing factors at each fixed time point. The fixed effect of adding this section controls for non-trending macro factors at each point in time . For example, due to exogenous shocks such as the trade war, China's overall economic growth rate slowed down in 2020, and this impact may have been greatly improved after the partial elimination of Sino-US tariffs in 2022. For such influencing factors that are not explicitly controlled in this paper, they can be mitigated by controlling for the temporal fixation effect.

2、**Use the asymptotic DID model:** General double difference models require that different individuals receive the same intervention time (e.g., all counties began to implement high-quality classification assessments in 2018) to construct multiple difference estimates. However, as already pointed out in the introduction, in fact, the change of county assessment is different in different provinces, and the timing of the start of different counties is also different, and even if the policy has already begun to be implemented, the real effect may be gradual. Therefore, this paper first constructs variables to represent the time that has elapsed since the policy $time_{it}$ went into effect for the $i$ county. For example, if the policy takes effect in 2018, then 2 will be taken in 2020 $time_{it}$. Next, before $trend_{it} = treat_{it} \times time$ the policy is implemented, all $trend_{it}$ are set to 0; The value for the future $trend_{it}$ depends on the time that has elapsed since the policy took place.

3、**Add covariates:** Covariates contain factors that do not directly affect outcome variables through policy variables $Y$ . The double difference method assumes that policy variables are

systematically different between the treatment and control groups, but if covariates are not added, it may lead to **omitted variable bias,** which further affects the regression results. In particular, it should be noted that if the influence of covariates on the outcome variables is not systematic, then the fixed effect already controls this part of the influencing factors, and there is no need to add covariates; The inclusion of model covariates should have a trending effect on the outcome variable. Therefore. All the covariates in this article add trends by multiplying with time.

4、**Use propensity score matching (PSM): Propensity score matching refers to the** selection of the most relevant control group individuals for each treatment group individual based on observable variables for score matching, and the treatment effect is estimated by two pairs and averaging the effect. It should be noted that the use of propensity score matching can only control the endogenous problem caused by observable variables, which is essentially equivalent to effectively compressing the information.

First, focus on the changes in the county's economic development in the multi-mission tournament mode. The processing variables defined here $\text{treat}_{it} = \text{treatment\_economy}_{it}$ examine the logarithmic value of county GDP (i.e., rate of change), county GDP per capita, and county GDP per capita difference. Here, multiple economic development indicators are used as proxy variables for economic development, and a robustness test is also indirectly constructed to verify the partial robustness of several proxy variables used in this paper and regression causal effects.

$$\log(gdp_{it}) = \alpha_i + \lambda_t + \beta_i treat_{it} + \gamma_i trend_{it} + \delta X_{it} + \epsilon_{it}$$
$$\log(pergdp_{it}) = \alpha_i + \lambda_t + \beta_i treat_{it} + \gamma_i trend_{it} + \delta X_{it} + \epsilon_{it}$$
$$gdp_{it} = \alpha_i + \lambda_t + \beta_i treat_{it} + \gamma_i trend_{it} + \delta X_{it} + \epsilon_{it}$$

Among them: is the multiplication of whether the intervention occurred and whether it is a dummy variable of the treatment group, $treat_{it}$ and the multiplication term of the time of the t year after the occurrence of the policy $trend_{it}$ and whether it is a treatment group, indicating the linear trend $t \in \{0,1,2,3,...\}$ of the policy.

TABLE 11 BENCHMARK REGRESSION: ECONOMIC DEVELOPMENT SPECIALIZATION LEVEL TEST, ORDINARY MULTI-TIME/MULTI-TIME ASYMPTOTIC DID MODEL

| | (1) | (2) | (3) | (4) | (5) | (6) |
|---|---|---|---|---|---|---|
| The variable name | lgdp | pergdp | lpergdp | lgdp | pergdp | lpergdp |
| treat_economy | 0.070*** | 0.552*** | 0.107*** | 0.015 | 0.093 | 0.0261* |
| (whether to intervene or not) | (4.77) | (6.20) | (6.02) | (0.93) | (1.63) | (1.71) |
| trend_economy | | | | 0.006*** | 0.058*** | 0.004** |
| (Policy Trends) | | | | (3.69) | (7.03) | (2.76) |
| Constant | 13.188*** | 1.986*** | 0.261*** | 13.175*** | 1.820*** | 13.180*** |
| (constant term) | (10,354.39) | (259.04) | (170.43) | (3,016.12) | (76.76) | (3002.55) |
| Trend control variable | No | No | No | No | No | No |
| County fixation effect | Yes | Yes | Yes | Yes | Yes | Yes |
| Time fixation effect | Yes | Yes | Yes | Yes | Yes | Yes |
| Number of observations | 30,347 | 23,751 | 21,486 | 30,347 | 23,751 | 21,486 |
| Adjust the R side | 0.977 | 0.817 | 0.959 | 0.977 | 0.817 | 0.959 |

Note: (1) Robust clustering t-statistic is shown in parentheses, clustered according to individual and time. (2) p<0.01, ** p<0.05, * p<0.1

The results of the benchmark regression (1)(2)(3) showed that whether using the logarithmic value of total GDP or the logarithmic value of GDP per capita or per capita GDP, the economic development level of the post-processing group classified as economic counties was significantly higher than that of the control group (P<0.01). (4)(5)(6) is a further split of the previous set of benchmark regressions, and in the case of asymptotic trend terms, is used to analyze whether the policy effect is due to immediate or gradual occurrence in the future trend. The regression results show that the estimated coefficients in the model are all positive, $trend_{it}$ indicating that the future trend can better reflect the policy effect.

In the extended form of regression, the modification of the causal effect estimation by adding multiple covariates is considered. In addition to the variables already estimated by the benchmark model, there are still multiple exogenous variables that may influence economic development trends. Because these trends may be linear, they cannot be absorbed by individual fixed effects or temporal fixed effects. This paper mainly considers the changes of policy treatment effects under covariates such as latitude and longitude, altitude, whether it is the boundary of a province, and whether it is a poor county. The results of (7), (8), (9) showed that the trend effect of the treatment group was still significant in several variables (P<0.01). In particular, (10), (11), (12) adds the square and cubic multiplication term of covariate and time, To further control for the nonlinear trend of the covariate, it is found that for the three forms of the result variable Y, the results of regression are still statistically significant (P<0.01) All of

the above results can be explained: **after being classified as an economic county, the total GDP growth rate, The per capita GDP and per capita GDP growth rates have increased significantly.**

TABLE12 EXTENDED REGRESSION: TEST OF ECONOMIC DEVELOPMENT SPECIALIZATION LEVEL AFTER ADDING COVARIATES, MULTI-TEMPORAL ASYMPTOTIC DID MODEL

|  | (7) | (8) | (9) | (10) | (11) | (12) |
|---|---|---|---|---|---|---|
| The variable name | lgdp | pergdp | lpergdp | lgdp | pergdp | lpergdp |
| treatment_trend_economy (Policy Trends) | 0.009*** | 0.049*** | 0.014*** | 0.009*** | 0.046*** | 0.013*** |
|  | (5.33) | (6.09) | (7.81) | (5.15) | (5.75) | (7.04) |
| post_ce_economy (whether to intervene or not) | 0.017 | 0.087 | 0.129*** | 0.028** | 0.143*** | 0.020 |
|  | (1.12) | (1.55) | (7.34) | (1.98) | (2.71) | (1.23) |
| The first-order trend of the control variable | Yes | Yes | Yes | Yes | Yes | Yes |
| The second-order trend of the control variable | No | No | No | Yes | Yes | Yes |
| Controls the third-order trend of the variable | No | No | No | Yes | Yes | Yes |
| Time fixation effect | Yes | Yes | Yes | Yes | Yes | Yes |
| County fixation effect | Yes | Yes | Yes | Yes | Yes | Yes |
| Observations | 30,347 | 23,751 | 23751 | 30,347 | 23,751 | 23751 |
| R-squared (adjust the R side). | 0.978 | 0.870 | 0.963 | 0.979 | 0.873 | 0.963 |

Note: Robust clustering t-statistic is shown in parentheses, clustered by individual and time. p<0.01, ** p<0.05, * p<0.1

Next, consider the improvement of the professional level of ecological environment development before and after the policy. As with the examination of the degree of specialization of economic development, this paper considers the differential value of PM 2.5 in the county and the logarithmic value of PM 2.5 as proxy variables for ecological environment development, and uses latitude and longitude, altitude, whether it is a provincial border, whether it is a poor county, etc Covariates. Although in the baseline model (1)(2), no direct basis for a significant reduction in pollution levels was identified before and after the intervention (P<0.05). ); However, after further splitting the model, it was found that the trend term part of the model was significantly negative. After controlling for more variables, although the tendency of the model's estimated treatment effect was weakened, the direction remained negative and significant at the 9to 5% confidence interval. In general, the PM2.5 growth rate of counties has indeed decreased after being classified as ecological counties compared with those not classified as ecological counties, but this reduction trend may have a certain lag, and it does not begin to change in the year of the policy.

$$\Delta pm25_{it} = \alpha_i + \lambda_t + \beta treat_{it} + \gamma_i trend_{it} + \delta X_{it} + \epsilon_{it}$$

$$\log(pm25_{it}) = \alpha_i + \lambda_t + \beta treat_{it} + \gamma_i trend_{it} + \delta X_{it} + \epsilon_{it}$$

TABLE 13 EXTENDED REGRESSION: TEST OF SPECIALIZATION LEVEL OF ECOLOGICAL DEVELOPMENT AFTER ADDING COVARIATES, MULTI-TEMPORAL ASYMPTOTIC DID MODEL

|  | (13) | (14) | (15) | (16) | (17) | (18) |
|---|---|---|---|---|---|---|
| The variable name | lpm25 | dpm25 | lpm25 | dpm25 | lpm25 | dpm25 |
| post_ce_ecology | -0.018* | 0.006 | 0.036*** | -0.029*** | 0.016 | -0.016*** |
| (whether to intervene or not) | (-1.72) | (1.57) | (3.44) | (-6.56) | (1.63) | (-3.21) |
| treatment_trend_ecology |  |  | -0.006*** | 0.002*** | -0.002** | -0.001** |
| (Policy Trends) |  |  | (-5.99) | (9.25) | (-2.29) | (-2.33) |
| Control variables (up to the third order trend). | No | No | No | No | No | No |
| Time fixation effect | Yes | Yes | Yes | Yes | Yes | Yes |
| County fixation effect | Yes | Yes | Yes | Yes | Yes | Yes |
| Observations | 34,001 | 32,210 | 34,001 | 32,210 | 34,001 | 32,210 |
| R-squared (adjust the R side). | 0.927 | 0.171 | 0.928 | 0.172 | 0.935 | 0.183 |

Note: Robust clustering t-statistic is shown in parentheses, clustered by individual and time. p<0.01, ** p<0.05, * p<0.1

Finally, in order to avoid the sample bias caused by the model used and further control the endogeneity of observable variables, the DID-PSM method is used to improve the model. Often people want to evaluate the effects of a public policy after implementation, and for this purpose, "treatment groups" and "control groups" need to be constructed to evaluate the treatment effect. However, the data are usually from non-randomized observational studies, and the initial conditions of the treatment group and the control group are not exactly the same, so there is a selection bias problem. The propensity score matching (PSM) method uses the propensity score function to compress the information of a multidimensional vector into one dimension, and then match it based on the propensity score. In this way, under the given observable characteristic variables, the individuals in the treatment group and the individuals in the control group are as similar as possible, thus alleviating the problem of selection bias of treatment effects.

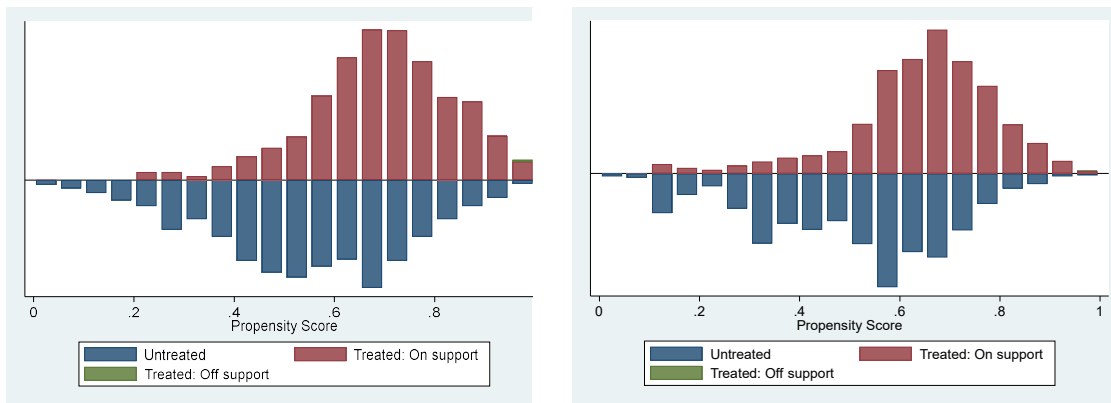

(A) ECONOMIC GROUP MATCHING EQUILIBRIUM    (B) ECOGROUP MATCHING EQUILIBRIUM

FIGURE 3 EQUALIZATION TEST FOR PROPENSITY SCORE MATCHING

| Regression group | Eco group | | Economic group | |
|---|---|---|---|---|
| | (19) | (20) | (21) | (22) |
| | OLS | PSM-OLS | OLS | PSM-OLS |
| post_ce_ecology (whether to intervene or not) | -1.891** (-3.05) | -3.162*** (-4.52) | 0.0702** (2.92) | 0.0854** (2.63) |
| treatment_trend_ecology (Policy Trends) | -0.692** (-3.08) | -1.301*** (-4.99) | -0.0155 (-1.54) | -0.0153 (-1.36) |
| Time fixation effect | Yes | Yes | Yes | Yes |
| County fixation effect | Yes | Yes | Yes | Yes |
| Control variables | Yes | Yes | Yes | Yes |
| N (number of groups). | 3384 | 3379 | 2299 | 2283 |
| adj. R-sq (robust R square). | -0.114 | -0.012 | 0.203 | 0.343 |

Note: Robust clustering t-statistic is shown in parentheses, clustered by individual and time. p<0.01, ** p<0.05, * p<0.1

The regression results after the propensity score match show that by alleviating the selectivity bias of the observable variables in the model, the policy treatment effect of both the economic group and the ecological group is partially improved. Changing the model setting did not have a significant effect on the direction of causal effects.

*E. Refute: Common trends, policy exclusivity, and other placebo tests*

Rebuttal refers to a robustness test for the causal effects of the model's estimates. When refuting the DiD model, three tests are mainly considered: **the common trend** test**, the policy exclusivity** test **and the placebo test.**

**In the common trend test, this paper adopts Callaway and Anna (2020)** method for processing group weight correction **for staggered DiD at intermediate time points.** The specific test method is to include the t-period advance of the multiple statistic in the double difference model, and if there is a common trend, then the coefficient multiple statistic in the previous t period should not be significant; And because the policy brings a significant doubling statistic for the year of policy implementation . To ensure that the estimated placebo effect (years before policy implementation) is comparable to treatment effect (years after policy implementation), the year prior to policy implementation was chosen as a reference.

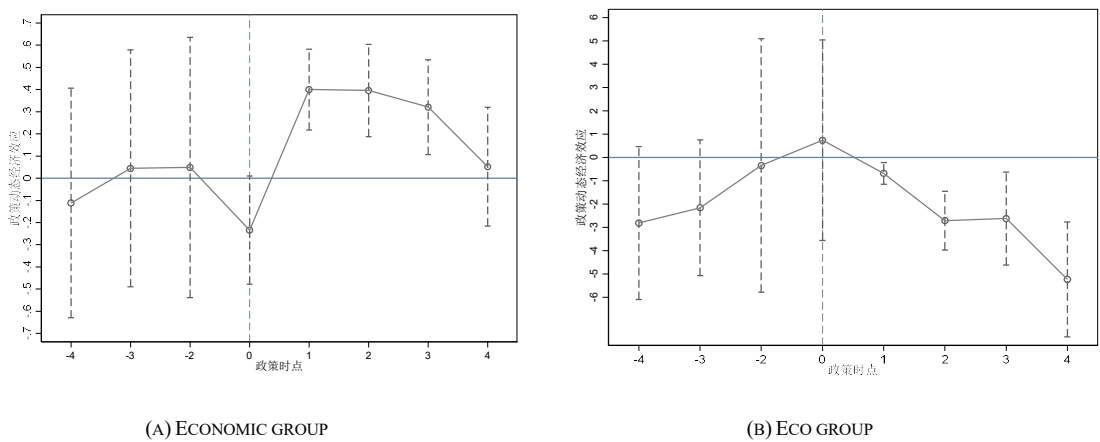

(A) ECONOMIC GROUP　　　　　　　　　　　　　(B) ECO GROUP
FIGURE4 COMMON TREND TEST: TIME PLOT OF POLICY TREATMENT EFFECTS AT VARIOUS LEVELS

**The test results of** the common trend of economic specialization **show that the common trend exists, and the policy effect is significant but not sustainable.** Specifically, the logarithmic value of county GDP was used as a representative outcome variable in the common trend test of economic specialization, and it was found that the placebo year economic development before the policy implementation was not significantly different from 0 ($P< 0.05$). , i.e., the treatment and control groups were similar in terms of economic development trends before policy implementation; In the first and second years of policy treatment, GDP growth in the treatment group was significantly higher than in the control group, and this effect reached a maximum in the second year; Subsequently, at the third and fourth years of treatment initiation, the estimated treatment effect decreased, by which time the treatment effect was no longer significant.

TABLE14 COMMON TREND TEST: TREATMENT EFFECTS OF ECONOMIC GROUPS

| Hysteresis/advance order | Estimators | Standard error | 95% lower limit | 95% cap | Sample size | The amount of conversion |
|---|---|---|---|---|---|---|
| $Placebo_1$ | 0.009 | 0.01 | -0.002 | 0.020 | 9604 | 396 |
| $Placebo_2$ | 0.009 | 0.01 | -0.006 | 0.025 | 8173 | 395 |
| $Placebo_3$ | 0.010 | 0.01 | -0.009 | 0.029 | 7671 | 395 |
| $Effect_0$ | 0.018 | 0.01 | 0.006 | 0.030 | 10992 | 412 |
| $Effect_1$ | 0.025 | 0.01 | 0.010 | 0.041 | 9513 | 405 |
| Effect$_2$ | 0.017 | 0.01 | -0.007 | 0.042 | 8915 | 402 |
| Effect$_3$ | 0.010 | 0.02 | -0.021 | 0.040 | 7565 | 374 |
| Effect$_4$ | 0.020 | 0.01 | -0.007 | 0.048 | 6411 | 321 |

In the test of the common trend of ecological environment specialization, the logarithmic value of PM2.5 was used as a representative outcome variable. In terms of results, there was a common trend in both the fourth and third years of policy implementation in the placebo years before policy implementation. Although there was no common trend in the second year before the implementation of the policy, the direction was opposite to the direction verified in this paper, that is, the individuals in the pre-treatment group had even higher PM2.5 values than the non-treatment group. No significant treatment effect was directly observed in the year when the implementation of the policy lagged behind, but from the first year after treatment, the mean value in the following years was negative. And from the overall trend, there is a clear downward trend compared with before treatment. This not only shows the effect of policy treatment, but also proves that there is a lag in the response of ecological environment governance to policy promulgation in the above model. This lag may be caused by the need for a certain amount of time for industrial transformation and upgrading, factory transformation, and ecological infrastructure construction.

TABLE15 COMMON TREND TEST: TREATMENT EFFECTS OF ECOLOGICAL GROUPS AT EACH STAGE

| Hysteresis/advance order | Estimators | Standard error | 95% lower limit | 95% cap | Sample size | The amount of conversion |
|---|---|---|---|---|---|---|
| Placebo$_1$ | 0.06 | 0.016 | 0.03 | 0.09 | 12930 | 370 |
| Placebo$_2$ | 0.01 | 0.008 | 0.00 | 0.03 | 11248 | 338 |
| Placebo$_3$ | -0.01 | 0.011 | -0.03 | 0.01 | 11036 | 338 |
| Effect$_0$ | 0.05 | 0.010 | 0.03 | 0.07 | 14721 | 389 |
| Effect$_1$ | -0.07 | 0.007 | -0.08 | -0.06 | 13041 | 357 |
| Effect$_2$ | 0.00 | 0.011 | -0.03 | 0.02 | 12829 | 357 |
| Effect$_3$ | -0.05 | 0.014 | -0.08 | -0.03 | 11285 | 314 |
| Effect$_4$ | -0.06 | 0.013 | -0.08 | -0.03 | 9714 | 301 |

In the estimation of this model, there may also be other policy effects, which make the policy treatment effect in this paper overestimated or underestimated. Therefore, in the exclusion test, in order to confirm that the impact of the change of county assessment form on the degree of local specialization is not a pseudo-causal effect, it is expected that the interaction term between dummy variables and multiple estimators in the year in which other policies take effect is added to re-estimate the model. If the estimation coefficient of the doubling estimator is large after adding the new interaction term, it may indicate that the estimation of the model is not robust; However, if the difference between the doubling estimators is not large compared with before, it can be said that the treatment effect estimated in this paper is robust. Around 2013, China issued a series of relevant documents on ecological construction and economic development, such as the Ministry of Environmental Protection of China, which officially implemented the National Environmental Protection Regulations and Environmental and Economic Policy Construction Plan of the Twelfth Five-Year Plan, which marked the official beginning of the national environmental and economic reform in the 12th Five-Year Plan. From the perspective of time, the promulgation year of this policy basically coincides with the division time of ecological counties and economic counties in the county economic assessment examined in this paper. Combined with the conclusion of Shi Daqian (2017), this paper uses the interaction term between dummy variables and doubling estimators around 2013 as control variables to further investigate the robustness of the model.

TABLE16 POLICY EXCLUSIVITY TEST: USING 2013 POLICY DUMMY VARIABLES

|  | (1) lgdp | (2) lpergdp | (3) dpm25 | (4) lpm25 |
| --- | --- | --- | --- | --- |
| DID | 0.130*** | 0.157*** | -0.016*** | -0.002*** |
|  | (0.019) | (0.026) | (0.004) | (0.006) |
| Policy implications after 2013 | -0.035* | -0.022 | 0.016*** | -0.041*** |
|  | (0.019) | (0.024) | (0.004) | (0.007) |
| Fixed effect | Yes | Yes | Yes | Yes |
| Control variables | Yes | Yes | Yes | Yes |
| N (sample size). | 30347 | 23751 | 32210 | 34001 |
| adj.$R^2$ | 0.980 | 0.963 | 0.161 | 0.937 |

Note: Robust clustering standard errors are shown in parentheses, clustered according to individuals and time. $p<0.01$, ** $p<0.05$, * $p<0.1$

The results of the exclusivity test show that for the economic specialization index, the policy impact after 2013 has no significant effect on the multiple estimator in the paper, and the interaction between the multiple estimator and the policy is not significant. For the ecological specialization index, although the policy impact coefficient after 2013 is significant, it has not changed the direction of the multiplier estimate, and the size of the policy effect has been slightly revised downward.

**Next, a randomized placebo test was performed.** In the traditional DID model, where all units have the same policy time, placebo testing requires a fixed number of units randomly selected from all units as the experimental group. However, this approach is no longer applicable when the policy time of each unit is different in multi-phase DID. The solution is to randomly select the sample period for each sample object as its policy time. For example, this paper provides the data of 30 provinces in China from 2000-20 21 , and in multiple phases of DID, it is necessary to randomly select a certain year from 2000-2018 for each of these 30 provinces as its policy time. The random sampling was carried out 500 times under the self-sampling method and the kernel density function plot was plotted, and the results could not reject the null hypothesis that the policy treatment effect of the random policy was greater or less than 0.

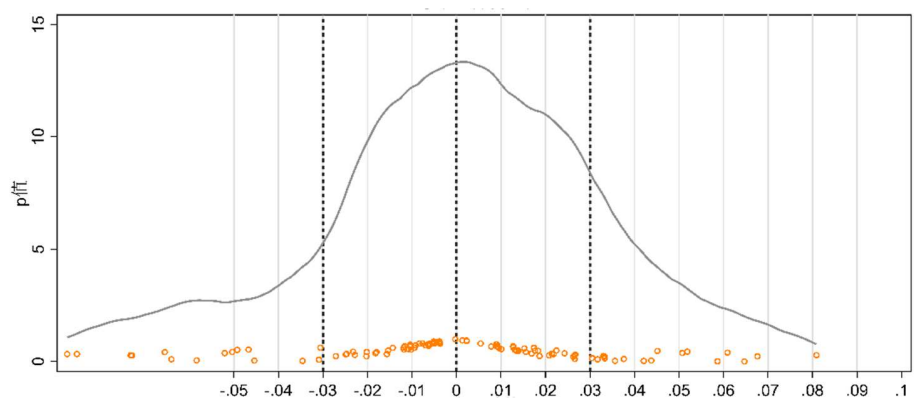

FIGURE5 PLACEBO TEST: KERNEL DENSITY PLOT OF RANDOMLY ASSIGNED SAMPLES

*F. Mechanism: Crowding out or compensating effect of effort level?*

In the estimation of the above model, the improvement of specialization in economic development or ecological environment development is tested respectively. When examining the level of specialization in economic development, the control group included both ecological counties and unclassified counties other than ecological counties and economic counties. Similarly, when testing the level of specialization of ecological environment development, the control group included both economic counties and unclassified counties other than ecological counties and economic counties.

In this section, the control group is further split in order to determine whether the overall effort of local officials in the level of specialization is shifted from one dimension to another (crowding out effect), or whether the level of effort in one dimension is increased when the level of effort in other dimensions remains unchanged, resulting in an increase in the overall level of effort (compensatory effect). This is done by adding another set of policy-treated dummy variables to the model to the test, which this paper defines as a crowding out effect variable. If the variable coefficient of the crowding out effect is significant and the direction is opposite to the direction of the policy variable, it means that the mechanism supports the crowding out effect, and the overall level of effort of local officials has not improved; If the variable coefficients of the crowding out effect are not significant, the mechanism supports the compensatory effect, at which point the overall level of effort of local officials may increase.

TABLE17 TEST OF POLICY MECHANISMS: CROWDING OUT OR COMPENSATING EFFECTS

|  | (1) lgdp | (2) lpergdp | (3) dpm25 | (4) lpm25 |
| --- | --- | --- | --- | --- |
| DID | 0.103*** | 0.139*** | -0.005** | -0.006* |
|  | (0.013) | (0.017) | (0.003) | (0.008) |
| Extrusion effect | 0.021 | 0.007 | 0.010*** | 0.017*** |
|  | (0.014) | (0.017) | (0.003) | (0.006) |
| Fixed effect | Yes | Yes | Yes | Yes |
| Control variables and temporal trends | Yes | Yes | Yes | Yes |
| N (sample size). | 30347 | 23751 | 32210 | 34001 |
| adj. R 2 (adjust the R side). | 0.980 | 0.963 | 0.161 | 0.937 |

Note: Robust clustering standard errors are shown in parentheses, clustered according to individuals and time. $p<0.01$, ** $p<0.05$, * $p<0.1$

The results show that there is no significant crowding-out effect observed in the indicators of economic specialization, but a more obvious crowding-out effect is observed in the indicators of ecological professional degree. Crowding-out effects are observed most clearly on ecological groups, since increases in eco-environmental effort are often accompanied by short-term economic sacrifices (eg, dismantling highly polluting factories). This also confirms the conclusion in the model of this paper from another aspect: that is, the essence of the improvement of the level of specialization is that competitors transfer part of their efforts from tasks that do not have a comparative advantage to tasks that have a comparative advantage. However, since the actual level of effort is unobservable, the proxy variables are not directly comparable in the two dimensions of ecology and economy, so it is impossible to directly verify that the aggregate effect does not exist, that is, the overall level of effort has not increased.

## D. References